\documentstyle[12pt,aasms4]{article}

\begin{document}

\title{A Near-Infrared Stellar Census of the \\
Blue Compact Dwarf Galaxy VII~Zw~403
\footnotemark[1]}

\author{Regina E. Schulte-Ladbeck}
\affil{University of Pittsburgh, Pittsburgh, PA 15260}
\authoremail{rsl@phyast.pitt.edu}
\author{Ulrich Hopp}
\affil{Universit\"{a}tssternwarte M\"{u}nchen, M\"{u}nchen, FRG} 
\authoremail{hopp@usm.uni-muenchen.de}
\author{Laura Greggio}
\affil{Osservatorio Astronomico di Bologna, Bologna, Italy, and
Universit\"{a}tssternwarte M\"{u}nchen, M\"{u}nchen, FRG}
\authoremail{greggio@usm.uni-muenchen.de}
\author{Mary M. Crone}
\affil{Skidmore College, Saratoga Springs, NY 12866}
\authoremail{mcrone@skidmore.edu}

\footnotetext[1]{Based on observations made with the NASA/ESA Hubble Space
Telescope, obtained at the Space Telescope Science Institute, which is operated
by the Association of Universities for Research in Astronomy, Inc., under
NASA contract NAS 5-26555.}

\begin{abstract}
We present near-infrared single-star photometry for 
the low-metallicity Blue Compact Dwarf galaxy VII~Zw~403. 
We achieve limiting magnitudes of F110W~$\approx$~25.5 and
F160W~$\approx$~24.5 using one of the NICMOS cameras
with the HST equivalents of the ground-based J and H filters. 
The data have a high photometric 
precision (0.1~mag) and are $>95$\% complete down to magnitudes of about 23, far
deeper than previous ground-based studies in the near-IR. The color-magnitude
diagram contains about 1000 point sources.
We provide a preliminary transformation of the near-IR photometry into the 
ground system. 

We investigate the tip-of-the-red-giant-branch method in the J and H bands to provide 
an empirical distance calibration. Combining our result with globular cluster data 
as well as stellar-evolution models, we recommend M$_{H, TRGB}$~=~-5.5($\pm$0.1) 
for -2.3$<$[Fe/H]$<$-1.5. We proceed to discuss the stellar content of VII~Zw~403 
using evolutionary tracks as well as a classification scheme based on 
optical and near-IR colors,
and comment on the detection of asymptotic giant branch stars and 
the Blue Hertzprung Gap. We use M$_H$ as an indicator of 
M$_{bol}$ for red stars after evaluating BC$_H$ at low metallicity. 

We calculate the fractional contribution of individual
stars from our color-magnitude diagram to the integrated light
of VII~Zw~403 and determine
which red stellar population dominates the integrated colors. 
We find that young red supergiants, and young and intermediate-age 
asymptotic giants, together provide about 50\% of the light in
I, J and H bands, whereas the old red giant stars
contribute less than 15\%.  Young, main-sequence stars 
and blue supergiants account for the remaining light and dominate in V. 
This explains the difficulties in discerning the nature of 
Blue Compact Dwarf galaxies when only integrated photometry is available.

\end{abstract}


\section{Introduction}

Blue Compact Dwarf galaxies (BCDs) are defined by their low luminosities 
(M$_B$${\ge}$-18), blue spectra with narrow emission lines,
small optical sizes, large H-I mass fractions (Thuan \& Martin 1981), and 
low oxygen abundances in their ionized interstellar gas
(e.g., Izotov \& Thuan 1999).  BCDs are known locally 
(z~$\approx$~0.02~--~0.03), with a few as far as 
z~$\approx$~0.1 (Thuan et al. 1994).
They are among the most vigorously 
star-forming dwarfs in the nearby Universe. 

The study of stellar populations in BCDs is interesting for
several reasons. First, ever since their discovery, it has been
an open question whether BCDs are ``young" galaxies which are forming
their dominant stellar population at the present epoch, or
``old" galaxies which formed
their first generation of stars earlier 
and are currently rejuventated by a starburst 
(Searle \& Sargent 1972). What is the nature of the BCDs?

Second, dwarf galaxies are the most common kind of galaxy at the
present epoch (e.g. Marzke \& da Costa 1997), and they may have been
even more numerous in the past (Ellis 1997). Guzm\'{a}n et al. (1998) 
propose that the (type CNELG~=~Compact Narrow Emission Line Galaxy) 
faint blue galaxies at redshifts
of z~$\approx$~0.5 could be luminous BCDs that
are experiencing a strong starburst. The ensuing supernova explosions 
are hypothesized to blow out the entire gas supply, resulting in a rapid fading 
through passive evolution (e.g., Dekel \& Silk 1986, Babul \& Ferguson 1996). However, HDF-N 
observations do not see the large numbers of faint red dwarf galaxy
remnants predicted by this scenario (Ferguson \& Babul
1998). What happened to the faint blue excess in the last few Gyr?

This paper addresses the star-formation history (SFH) of BCDs.
The stellar content of BCDs provides a fossil record of their SFHs. 
However, most BCDs are at such large distances
that only their most luminous stars can be resolved, even with
the HST. 

VII~Zw~403 is a very nearby (4.5~Mpc) example of the BCD class. It
is a key object to study with HST since it is close enough to be well resolved into
individual stars (Schulte-Ladbeck et al. 1998, hereafter SCH98; Lynds et al. 1998). 
VII~Zw~403 is a type ``iE" BCD in the classification of Loose \& Thuan (1986).
This is by far the most common type of BCD, and is considered 
characteristic of the BCD phenomenon. The outer isophotes of iEs 
are elliptical, whereas
the brightest star-forming regions are distributed somewhat irregularly
in the vicinity of the center --- although not at the exact center. This 
description is strikingly similar to that of 
early-type (dSph, dIrr/dSph) galaxies in the Local Group (see Mateo 1998). 
VII~Zw~403 exhibits a smooth, elliptical background sheet 
(see the R-band image of Hopp \& Schulte-Ladbeck
1995) with an integrated color consistent with an old and metal poor
stellar population (Schulte-Ladbeck \& Hopp 1998). 
The background sheet resolves in
HST/WFPC2 images into individual red giant stars. The scale length
derived from the resolved stars is about 50\% larger
than that of the large spheroidals of the Local Group. 
Schulte-Ladbeck et al. (1999, hereafter SHCG99) propose
that VII~Zw~403 posseses an old underlying stellar population. This supports
earlier suggestions 
that all BCDs showing extended halos of red color might
be old galaxies. It also strengthens the possible evolutionary link
between BCDs and early-type dwarfs (e.g., Sung et al. 1998). 
VII~Zw~403 has a present-day metallicity of about Z$_\odot$/20 (see SHGC99),
a moderate star-forming rate of 0.013~M$_\odot$yr$^{-1}$ (Lynds et al. 1998), 
a large H-I mass of about 7x10$^7$~M$_\odot$ (SHCG99), 
an extended H-I envelope (Thuan 1999, private communication), and  
a large outflow of hot gas detected in X-rays
(Papaderos et al. 1994). Therefore, VII~Zw~403 is clearly in the process
of cycling gas in and out of the deepest part of its gravitational potential; 
this may regulate its star-formation rate and promote transitions 
in optical morphology between early and late type over Gyr time-scales.

VII~Zw~403 has an H-I derived heliocentric velocity
of -92~km/s (Tully et al. 1981). Tully et al. combined this with the
fact that VII~Zw~403 is somewhat resolved in their ground-based
images to associate it with the M81 group, at an assumed distance 
of 3.25~Mpc. 
When we applied the tip-of-the-red-giant-branch (TRGB) method to derive 
its distance from the I-band luminosity function of the halo
stars, we found that VII~Zw~403 is about 40\% further away, 
at about 4.5~Mpc (SHCG99). A survey of several emission-line galaxy catalogs
(e.g., Kunth \& S\`{e}vre 1986, Thuan \& Martin 1991, Terlevich et al. 1991, 
Salzer et al. 1995, Pustil'nik et al. 1995, Popescu et al. 1996 and references
therein) reveals additional BCDs with small positive recession 
velocities corresponding to the 5 to 10~Mpc distance range. We selected
another four well studied BCD/dIrrs for NICMOS observation, and these will
be discussed in future papers (Hopp et al. 2000).

Near-IR observations have already been used to investigate 
the nature of BCDs. Tully et al. (1981) observed VII~Zw~403 in J and H
in an aperture centered on the starburst, and attributed the near-IR
flux to red supergiants. Thuan (1983, 1985) 
obtained integrated, near-IR photometry of a large sample of BCDs 
and argued that he had found an
old population of K and M giants. Unfortunately, as Thuan (1983) recognized, it
is difficult to discriminate between a population of young red
supergiants and one of old red giants using only the total optical/infrared
colors of a star-forming galaxy, because of the overlap in the effective
temperature ranges of red supergiants and red giants. 
Campbell \& Terlevich (1984) obtained photometric
CO indices of BCDs and asserted that the population detected in the 
near-IR is primarily composed of supergiants from the current starburst. 
Subsequent near-IR imaging has been used to resolve
the more centrally concentrated star-forming regions which are dominated by
the younger supergiants,
from the more extended background sheets, which are potentially dominated 
by older red giants  
(e.g., James 1994, Vanzi 1997, Davies et al. 1998). However,
James (1994) found that intermediate-age, asymptotic giant branch
(AGB) stars could be responsible for as much as 50\% of the near-IR
emission of some BCDs, further complicating the interpretation of
near-IR data in terms of stellar populations. Whereas integrated 
colors and color profiles of mixed-age stellar populations
yield ambiguous results, it is in principle possible to distinguish among 
contributions from young red supergiants (RSGs), intermediate-age
AGB stars, and stars on the first ascent red giant branch (RGB) 
with the help of color-magnitude diagrams (CMDs). 

There is little work in the literature regarding resolved
stellar populations of similar star-forming galaxies in the near-infrared. The local
starburst cluster R~136 in the 30~Dor region of the LMC has 
been studied with adaptive optics in the near-IR, but there is
no old population (Brandl et al. 1996). NIC2 results on R~136 (Walborn
et al. 1999) focus on the young, and still embedded, massive stars. 
Adaptive optics, near-IR photometry has not yet been applied successfully to the study of 
stellar populations in the more distant star-forming galaxies of the Local Group 
(e.g., Bedding et al. 1997). The DeNIS survey which is currently in progress at ESO, 
is mapping the Magellanic Clouds (MC) simultaneously in I, J, 
and K$_s$, to limiting magnitudes of
about 18, 16 and 14, respectively. DeNIS data clearly reach below
the TRGB, as evidenced by preliminary color-luminosity diagrams and luminosity functions
published by Cioni et al. (1999). In addition, the CMDs are well populated
with blue stars, presumably the upper main sequence and blue supergiants.  
Recently, the post-starburst
dIrr IC~10 and the dIrr IC~1613 were resolved in J and H with limiting magnitudes of
J~$\approx$~18 and H~$\approx$~17.5 (Borissova et al. 1999). 
These data are considered deep enough to show the TRGB (m-M = 24.0 and 24.2, 
or distances of a few hundred Kpc, respectively). However, inspection
of their luminosity functions shows the suspected TRGB occurs at the
very limit of their data where incompleteness due to crowding is 
a severe problem. The dIrr galaxy NGC~3109, 
at a distance of m-M = 25.6 or 1.4~Mpc (Alonso et el. 1999), is located
in the outer regions of the Local Group. Alonso et al. observed it in the
near-IR to limiting magnitudes of about 20 and 19 in J and H.
The RGB is below the detection limits in the near-IR.

There are no known BCDs in the Local Group.
Much deeper limiting magnitudes and better spatial resolution 
are required if we wish to resolve the old stars in galaxies at  
distances of up to 10~Mpc (m-M = 30.0). In order to reach the 
TRGB in such BCDs, we can make use of recent improvements in near-IR
imaging. The peak of the spectral energy 
distribution of RGB stars occurs in the 
near-IR. A K5 giant, for example, has colors of V-I = 2.1 (Johnson 1966) 
and V-H = 3.5 (Koornneef 1983), suggesting that a significant gain of near-IR
over optical imaging is possible. Near-IR observations 
are therefore a promising route to mining the
old stellar populations in BCDs at large distances. The NICMOS instrument 
aboard HST offered the opportunity to study
stellar populations in the near-IR at high spatial resolution and
deep limiting magnitudes. VII~Zw~403,
for which optical and near-IR single-star photometry is discussed
in this paper, provides the ``proof of concept." Hopp et al. (2000) 
give a  preview of our
intended applications to the additional four galaxies observed by us
with HST/NICMOS. 
 
Here we present near-IR images of VII~Zw~403 
obtained with HST's 
NIC2 camera. The galaxy is resolved into single stars in the near-IR,
several magnitudes deeper than previously achieved for dIrrs
from the ground, and sufficiently deep to yield useful measurements of 
stellar magnitudes and colors. We compare these measurements with 
WFPC2 observations of the same galaxy, to form 
an empirical characterization of the stellar content 
in the near-IR. We measure the TRGB in the HST equivalents of the J and H bands,
and provide a calibration of the TRGB method. In future
publications, we will apply these TRGB fiducials to NICMOS observations 
of other resolved BCD and dIrr galaxies for which there are no optical 
TRGB data. We compute the fractional light that different
stellar types contribute to the integrated colors in optical and near-IR
bands. Finally, we comment on the nature of BCDs.

\section{Observations and reductions}

The star-forming regions of VII~Zw~403 are located near 
the center of an elliptical
background-light distribution (Hopp \& Schulte-Ladbeck 1995, Schulte-Ladbeck
\& Hopp 1998). In the HST/WFPC2 observations with which we compare
our NICMOS photometry, the 
star-forming centers are situated in the PC chip. An image
of this region was published as Plate 1 of SCH98. The three WF chips 
cover part of the elliptical background-light distribution of VII~Zw~403, but,
due to the geometry of the arrangement of the WF chips, not the entire galaxy. 
A record of the UV/optical imaging obtained with the WFPC2 on 1995 July
7 in F336W, F555W, F814W, and F656N filters, the equivalent of
the U, V, I and H${\alpha}$ passbands, was given in SCH98. 
Errors and completeness
fractions for stellar photometry in the continuum bands, and comments
on the transformation into the Johnson-Cousins system can be found in
SHCG99.

\subsection{NICMOS imaging}

NICMOS observations of VII~Zw~403 were obtained on 1998 July 23 as part of 
GO program 7859. Information about the observations can be gleaned
directly from the STScI WWW pages linked to this program ID. 

The NICMOS instrument houses three cameras, the NIC1, NIC2 and NIC3,
 in a linear arrangement. Data for VII~Zw~403 were 
obtained with all three cameras operating simultaneously,
to mitigate some of the detector problems which became known during the
phase-II stage of the proposal. However, although all three detectors
were collecting photons, not all three yield data 
which are useful for this study.
The NIC3 cannot be brought into focus simultaneously with NIC1 and NIC2, 
and even NIC1 and NIC2 are not completely confocal. We conducted the
observations at the compromise focus position between NIC1 and NIC2. 
We performed all 
of our imaging observations
in the F110W and F160W filters, the rough equivalents of the J and H
bands. Although our primary goal was to obtain
deep imaging with both high resolution and large area coverage of regions located in 
``Baade's red sheet \footnotemark[2]", our observing strategy was designed to optimize the 
scientific return and minimize the risk of using
a new instrument with somewhat uncertain performance. We therefore chose to locate 
the NIC2 chip, which has a larger area and higher sensitivity at lower resolution 
than the NIC1 chip, within the central starburst region of the galaxy rather
than in the elliptical background sheet. In order not to constrain the
schedulability of the observations limited by the lifetime of the NICMOS cryogen,
we also decided not to request a specific orientation of the observations, and
so the NIC1 chip was positioned within the outskirts of the background 
sheet of VII~Zw~403 at that spacecraft roll which happened to occur on
the observing date. The geometry of the NICMOS observations relative to 
the WFPC2 observations can be gleaned by comparing Figure~1 to Plate~1 of SCH98. 
For reference, the NIC2 camera has a field of view of 19.2"x19.2" 
and a pixel scale of 0.075". Thus the area of the NIC2
is only about 30\% of the area of the PC.

\footnotetext[2]{The term ``Baade's red sheet" 
has its roots in Baade's foundation of the stellar population concept, i.e., the
definition of Population I and II.
An interesting historical perspective on his ideas was given by Sandage (1986). }

The observations were split into several exposures and read out in
MULTIACCUM mode. The parameters were chosen to address several issues 
which are laid out in the NICMOS 
documentation (detector read-out noise, saturation of bright sources, 
amplifier glow), and to optimally fill the total time 
available per HST orbit. From one exposure to
the next, the NICMOS camera was dithered in the X direction. This
procedure allowed for a better sampling of the PSF, better
flux measurements, and the removal
of cosmic rays and background in the processing of the data.

Due to its sensitivity and very small field of view, the NIC1 camera produced
images which contain very few point sources. We will
not discuss the NIC1 data in this paper. The NIC3 images exhibit several
bright stars, but the stellar images are out of focus. Therefore, we will
not discuss 
the NIC3 data in this paper either. 

In the F110W filter, we gathered a total of six individual exposures, 
three having integration times of 1023.95~sec each, and three of 767.96~sec. Severe 
cosmic-ray persistence in the images (see below) convinced us not to 
include one of the short exposures in the final dataset. The total 
integration time for the F110W image is therefore 4607.77~sec.
In the F160W filter, we took a total of five exposures, each with an
integration time of 767.96~sec. Again, one of the datasets was so
badly affected by cosmic-ray persistence that we chose not to
add it into the final image. The integration time of the F160W filter
image used here is therefore 3071.84~sec.

The NIC2 observations were reduced with the latest
reference files available in the calnica pipeline, and the individual
observations were combined into a mosaic with the calnicb pipeline.
However, this reduction was not satisfactory.
The data showed surprisingly few point sources compared to the
WFPC2 images. They displayed a high and
spatially non-uniform background. Aperture photometry
immediately indicated that the anticipated limiting magnitudes
were not achieved. In fact, initially the DAOPHOT (Stetson, Davis
\& Crabtree 1990) software did not
recognize many of the point sources visible by eye.

The observations of VII~Zw~403 are severly affected by what
are known as the ``pedestal" and the ``cosmic-ray persistence"
problems. Both of these problems result in the addition of
 spatially non-uniform, high backgrounds to the data, 
preventing the detection of faint stellar IR point sources
in our images. 

We are very grateful that Dr. M. Dickinson allowed us to use
his personal software to remove to a large extent the effect
of the pedestal-induced background in our data. To accomplish
removal of the pedestal, the raw data were first run through the
calnica pipeline without applying a flat-field correction.
We then used MD's software to interactively fit a pedestal-free 
sky level to the background in the
four quadrants of the NIC2 detector. The data were re-processed
through calnica with the flat-field flag on. The individual
exposures so reduced exhibited a much smoother and a lower 
background than the raw data. 

In order to avoid the cosmic-ray persistence,
we had requested that our data be obtained well away from
the South Atlantic Anomaly; however, this was not possible due
to the pressure on NICMOS time and the ensuing scheduling
difficulties during its short lifetime.
Because all of the exposures are affected by this 
problem at some level, the final data do not reach as deeply as
they would otherwise. 

We proceeded to combine the reduced data with calnicb.
However, careful inspection of the mosaiced images revealed
that the PSF of the point-sources varied noticably across
the images. Since DAOPHOT photometry is sensitive to
the shape of the PSF, we needed to minimize these PSF variations.
To do this, we combined our distortion-corrected exposures 
using a drizzling routine, instead of using the calnicb pipeline,
(which, at time of our data reduction, did
not take into account geometric distortion.) 
The drizzling process required a careful manual masking
of detector blemishes before the image combination.

The resulting images in F110W and F160W have a more uniform
and lower background, and a more uniform and rounder PSF, than the data
reduced by the pipeline. Owing to the dithering procedure, areas at the
edges of the images in X direction have a lower signal-to-noise,
which we trimmed off.

\subsection{Single-star photometry on NICMOS images}

The NIC2 single-star photometry is the result of an iterative process.
After the images were reduced as described above, we fitted
a PSF to about one hundred fairly isolated sources in each image
and carried out photometry. 
We set the zeropoint using the most recent 
photometric keywords available for the F110W and the F160W
filters used with NIC2.  We examined about a dozen sources in each
image to calculate the aperture correction
that we applied to the DAOPHOT PSF photometry. 
The CMD in F110W, F160W reached deeply enough  
to reveal the TRGB.
However, the photometric errors for the faint sources were large.

Therefore, we once again examined the background. The background
was still non-uniform due to imperfect pedestal correction, the remaining
cosmic-ray persistence, and bound-free and free-free emission
from ionized gas in the H-II regions. Smaller photometric errors
for the faint sources were obtained when the background was smoothed
with a 23x23 pixel median filter after all the identified 
point sources were removed
from the data. Photometry on this background yielded smaller photometric
errors for the stars that were previously identified by DAOPHOT, and
as a consequence, a tighter distribution of the stars on the CMD.
While an even more extended sea of very faint sources was visible on 
the images by eye before the background smoothing was applied, the 
high photometric errors on the sources that were identified with
DAOPHOT convinced us that photometry of these potentially real,
faint red giants was not feasible. The images which had the
smoothed background subtracted off are shown
in Figure 1.

When we overlayed the sources identified by DAOPHOT 
on the images, we noticed that the extended PSF,
especially in the F160W filter, yielded multiple identifications of
very bright sources. Possibly as a result of 
the pedestal removal and the drizzling, the
first diffraction ring was of a non-uniform brightness, and
DAOPHOT identified parts of the ring as 3-4 additional point
sources. These sources were usually several magnitudes below the
main peak and had a high error. However, more severly, we noticed
that in several cases the main core of the PSF of the brightest
stars was identified with two PSFs of equal brightness. After
much experimentation with the PSF and other clipping parameters
within DAOPHOT, we found that the best results were achieved
when the images were smoothed with a 3x3 pixel filter. 
In this way the sizes of the cores of the PSFs were degraded
slightly to a FWHM of 2.9~pixels (0.22") in F110W and
3.6~pixels (0.27") in F160W, but the brightness distribution
in the diffraction features was more uniform. An illustration
is provided in Fig.~2. We reapplied
DAOPHOT and the bright stars were 
found as single sources, while the sources erroneously
identified as faint stars within the PSFs of the bright stars
vanished. With this procedure we measured 2134 individual sources 
with residual errors smaller than 0.55 mag in F110W, and 1500 
sources in F160W. In constructing the CMD we required spatial
coincidence of the sources to within 3~pixels; 998 such sources
were found.

Finally, we re-investigated the photometry of the stars used
for the photometric calibration and the aperture correction.
By measuring these stars in a series of ever larger circular
apertures, we found that the PSF had not been encompassed
sufficiently by the 0.5" aperture. We calculated an
aperture correction to the 0.5" aperture (for which the
photometric conversion is given by the NICMOS photometric 
calibration) and re-applied the appropriate 
corrections to the data sets.
 
In Fig.~3, we display the internal errors of
the photometry. The errors for F110W exceed
0.1~mag for a magnitude of 23.7; for F160W 
they reach 0.1~mag at a magnitude of 22.5. We performed completeness
tests using the same procedure described in SHCG99. 
In Fig.~4, we summarize the results. 
The data are nearly complete ($>$95\% of
test stars recovered) for magnitudes $<$23; completeness
drops to 50\% at 25.4 in F110W and 24.1 in F160W.

\section{Results and Discussion}

The WFPC2 CMD in V and I contains 5459 sources, 
and is displayed as Fig.~5. Throughout most of this section, 
the [(V-I)$_o$, M$_{Io}$] CMD serves as a guide to interpreting 
the NICMOS results. Main-sequence (MS),
blue supergiant (BSG), blue-loop (BL), RSG, AGB and RGB stars
are all represented. This CMD is extensively 
discussed in SCH98, Lynds et al. (1998) and SHCG99; we have good knowledge of the 
location of various stellar phases on this diagram. 

Figure 5 shows the evolutionary tracks with Z=0.0004; Z=0.004 
(Z$_\odot$/50 and Z$_\odot$/5, respectively) from the
Padova library (Fagotto et al. 1994). These have been transformed 
into the observational (HST) plane by using bolometric corrections and 
color tables (Origlia \& Leitherer 2000)
produced by folding the HST filter/system response with model atmospheres from Bessell,
Castelli \& Plez (1998) with [M/H]~=~-1.5; [M/H]~=~-0.5. These transformation 
tables adopt  M$_{V,\odot}$~=~4.83 (e.g. M$_{bol,\odot}$~=~4.75, BC$_{V,\odot}$~=~-0.08)
and colors equal to zero for the model atmosphere representing
$\alpha$~Lyrae. The F555W and F814W magnitudes of these two grids were transformed 
to Johnson-Cousins
V and I magnitudes in the ground system using Table~10 of Holtzman et al. (1995), so
that the tracks are on the same photometric system as the VII~Zw~403
observations. 

In Fig.~5, we overlay a few tracks
of either metallicity onto the observations. The tracks nicely
illustrate the well-known age-metallicity degeneracy of the RGB
in broad-band colors, i.e., 
the tip of the first-ascent red giant branch of the Z=0.0004,
1~M$_\odot$ model which has an age of about 7~Gyr virtually coincides with that of the 
Z=0.004, 4~M$_\odot$ model which has an age of about 160~Myr, 
in the [(V-I)$_o$, M$_{Io}$] plane. Thus, additional
arguments based on the positional dependence of the RGB were 
employed in SHCG99 to suggest the presence of an old 
and metal-poor stellar population. 

A CMD of the near-IR photometry is displayed as Figure~6,
in terms of instrumental magnitudes in the Vega system.
A transformation of the F110W and F160W photometry into J and H 
is attempted below. Since additional systematic
errors are added in the process, we first discuss the F110W and 
F160W photometry. The foreground extinction towards VII~Zw~403 
(E(B-V)~=~0.025) is negligible at the central wavelengths
of the near-IR filters (computed using Cardelli, Clayton \&
Mathis 1989). Because VII~Zw~403 is situated at high
Galactic latitude, and the NIC2 camera has a very small size, the
contribution to the CMDs by Galactic foreground stars is also negligible. 

The NIC2 images were situated well within the PC images of VII~Zw~403.
We cross-identified sources 
found in both cameras by transforming the NIC2 coordinates into the
WFPC2 system. We then merged our photometry lists, and investigated
the distribution of stars on a variety of color-color diamgrams
and CMDs. There are 549 sources found in V, I, J, and H.

\subsection{Two-color diagrams and internal extinction}

We examined two-color diagrams (TCDs) for all combinations
of the data sets (Figure~7). 
The TCDs show two clumps of
stars corresponding to the blue and red plumes of the CMDs (see below), 
with only a few sources located outside of this main distribution
of stars. The reddening vectors
are approximately parallel to the distribution of stars even using
our seemingly advantageous long color baseline of V-F110W
vs. V-F160W. This can also be seen in Koornneef (1983), Fig. 1,
for ground-based V, J, H data. We do not discuss TCDs involving the U band 
here, due to the 
paucity of sources found in U and the near-IR bands.
 
We investigated the reason for observing a
few sources off the main clumps of stars in the TCDs, and found that 
the deviant data points can all be
explained by large photometric errors ($>$~0.1 mag) in one of three filters.
We thus attribute the sources which are offset from
the main distribution to measurement error and not to
internal reddening. This result of undetectable internal reddening 
in VII~Zw~403 is consistent with the seemingly unreddened location of
the blue plume in the [(V-I)$_o$, M$_{Io}$] CMD, centered on
0 mag, and with the fact that few stars change position from 
the red to the blue side when we plot CMDs with a larger 
color baseline (see below).

Furthermore, Lynds et al. (1998) derived E(B-V)~= 0.04--0.08 for
the internal extinction of stellar associations in VII~Zw~403. 
This corresponds to A$_V$~= 0.12--0.25,
and, using the Galactic extinction law of Cardelli, Clayton \& Mathis (1989),
translates into A$_J$~= 0.03--0.07 and A$_H$~= 0.02--0.05. As expected,
the extinction in the near-IR is very small.

\subsection{Luminosity functions in the near-IR and the TRGB method
for deriving distances}

Briefly, the TRGB method (Lee et al.
1993) makes use of the relative insensitivity to metallicty of
M$_{bol}$ of the tip of the first-ascent red giant branch, 
as well as the insignificance of line-blanketing
for the I magnitude of metal-poor red giants,
and the availability of well-calibrated bolometric corrections to
the I band based on the V-I color of the RGB. However, similar
calibrations for the F110W or F160W filters do not exist. Furthermore,
as illustrated in Fig.~8, while the absolute I magnitude at the TRGB is constant
below an [Fe/H] of about -0.7, those in J and H display
a metallicity dependence. In this section,
we discuss an empirical near-IR TRGB calibration based
on the VII~Zw~403 data, which is therefore valid for the VII~Zw~403
metallicity. In the following sections,
we will investigate the dependence of the near-IR TRGB
on metallicity both empirically and using models. 

In SHCG99, we derived the distance modulus of VII~Zw~403 to be (m-M)$_o$=28.23 
from the V-I color of the RGB for the halo population 
combined with the location 
of the TRGB in the I-band. This immediately allows
us to place an absolute magnitude scale on the near-IR CMDs
of VII~Zw~403, to identify the TRGB here, and to interpret
the stellar content of the near-IR CMDs. The RGB is clearly
distinguishable in the near-IR CMDs, as expected, as a
densely populated region at red colors and faint magnitudes, the
red tangle.

Comparing Fig.~5 with Fig.~6 we notice that, while showing great
morphological similarity, the near-IR CMDs do not 
exhibit the pronounced red tail of AGB stars
seen in the optical CMD. In the near-IR CMDs, the red plume continues into
the red tangle as a strikingly linear feature with colors
0.75$<$(F110W-F160W)$<$1.5. The RSG, AGB
and RGB stars are not well-separated in color, as we further illustrate
below. 
However, tracing along the red plume it is also evident that
a vast increase in the star counts occurs at faint magnitudes, 
and a TRGB can be distinguished from these data.

Luminosity functions in F110W and F160W were derived by counting stars in
0.1~mag bins in the color interval 0.75$<$(F110W-F160W)$<$1.5. 
As Fig.~9 shows, the luminosity functions display a sharp rise of the
star counts towards fainter magnitudes. We identify this rise with the TRGB. 
We emphasize 
that the TRGB occurs at apparent magnitudes
in F110W and F160W where the completeness of our data is still
very high (above 90\%). Using the above distance modulus,
we find the TRGB is located at 

\begin{center}
M$_{F110W, TRGB}$~=~-4.28~$\pm$~0.10~$\pm$~0.18  and\\
\smallskip
M$_{F160W, TRGB}$~=~-5.43~$\pm$~0.10~$\pm$~0.18, \\
\end{center}

\noindent where errors were computed 
from a combination of the random and systematic errors 
as discussed in SCH98. 

The near-IR TRGB values
are presumably valid at the metallicity of the halo population of 
VII~Zw~403, namely at $<$[Fe/H]$>$~=~-1.92$\pm$0.04 (Z$_\odot$/83). We compare
the empirical values with the 15~Gyr TRGB in our low-metallicity grid (evolution at
[Fe/H]=-1.7, atmospheres at [Fe/H]=-1.5, see above) and find that the models
yield M$_{F110W, TRGB}$~$\simeq$~-4.6 and M$_{F160W, TRGB}$~$\simeq$~-5.6,
in good agreement with the data.

These results provide one ``calibration" for the TRGB in the near-IR, and could
in principle be applied to other galaxies with old and similarly metal-poor 
giant branches. We now discuss 
the accuracy of the method by comparing our CMD in J and H to those of 
globular clusters (GCs).

\subsection{Transformation to J and H}

The STScI NICMOS team supplies on their home page 
data for five standard stars (release Nov. 25, 1998), including 
magnitudes in HST filters and standard ground-based filters. 
This small sample consists of a white dwarf, a G star, and three 
red stars of increasingly
redder color. The total color range of the stars is 
-0.04~$\le$~J-H $\le$~+2.08 and
the total magnitude range is 7.3~$\le$~H~$\le$ 12.7. 
The ground-based data of the NICMOS standard stars are mostly unpublished
(private communication from the NICMOS IDT to the STScI NICMOS team). 
At least one red star comes from the list
of Elias et al. (1982; CIT system of JHK magnitudes). 
The STScI NICMOS team mentions
discrepancies of the order of 0.1~mag between the ground-based 
data in their Table and 
those of Persson et al. (1998, LCO system; see this paper and 
references therein
for a comparison of the various ground-based systems). 
A complete transformation into a ground system is not yet available. 

We used the data of the NICMOS team to establish transformation equations
from the F110W, F160W magnitudes and color to the ground-based
system, applying a simple linear least-squares fit to the data:
\begin{center}
$ J-H~=~(0.035~\pm~0.14) + (0.758~\pm~0.061) * (F110W-F160W)$\\
\smallskip
$ H~=~(-0.001~\pm~0.06) + F160W - (0.091~\pm~0.027) * (F110W-F160W)$\\
\end{center}
The errors give the quality of the fit but of course do not
take into account any of the systematics (such as the offset mentioned
above, or the limited sampling of color range by the few standards). It is 
therefore difficult to estimate the
uncertainties introduced by applying these transformations. We have to
assume that the accuracy is no better than 0.1~mag, and probably worse.

In Fig.~10, we display the CMDs of VII~Zw~403 in J and H. Overlayed are the
tracks for the same stellar masses and metallicites as we used in Fig.~5.
The tracks were transformed to the ground-system using the above equations.
There is no major qualitative change in these CMDs as compared to those
in the instrumental system (cf., Fig.~6). Quantitatively, the main effect is 
a shift of the red plume towards the blue plume. 
An advantage of transforming the data
into the JHK system is that our CMDs and luminosity functions can be more 
readily compared with the few available ground-based data of similar galaxies in the near-IR 
(see Alonso et al. 1999, Borissova et al. 1999, Cioni et al. 1999). The transformation also 
enables a comparison of our data with ground-based photometry of GCs in the 
Milky Way and the LMC. We can also give the J, H magnitudes of the TRGB
of VII~Zw~403 (see Table~1).

In making comparisons of near-IR data sets, additional offsets may be encountered 
because the various ground-based observations were obtained in different realisations of
the JHK filter system. For a careful comparison, we need to transform all of the
data into the same system. 
Persson et al. (1998) extend  the bright near-IR standard star measurements of Elias et al. (1982) to
fainter magnitudes. They discuss how the LCO system compares to others
(UKIRT, see Casali \& Hawarden, 1992; CTI, see  Elias et al., 1982, AAO, see
Bessel \& Brett, 1988) and find rather good agreement between the UKIRT, 
CTI, and LCO systems in H and K. The J band is less straightforward. 
For instance, Elias et al. (1983) derive transformation formulae between the 
AAO system and their CTI system. Again, H and K show good agreement and the two systems can be
treated as identical systems for our purposes. 
However, a significant transformation
coefficient (about 0.8~mag) has to be taken into account for J.
This may be due to the large contribution of atmospheric features in this band.
In our discussion below, we shall
therefore focus our comparisons on the H-band results, where
the three systems discussed above can be assumed to be identical 
within the accuracy
of our transformation to the ground system.

\subsection{TRGB H magnitudes in GCs and stellar models} 

If we wish to use the VII~Zw~403-derived calibration 
as a distance indicator for other galaxies, we 
have to investigate the sensitivity to
metallicity of the TRGB luminosity in the near-IR.
To study this limitation of our estimator, we followed two approaches, using both
stellar evolution models and GC data for guidance.

First, we read off the magnitudes from the Padova
isochrones (Bertelli et al. 1994, see also Fagotto et al. 1994)
for 15 Gyr old stars for the largest available metallicity range. 
These data are shown in Fig.~8. Bertelli et al. provide their isochrones in the
Johnson-Cousins system in the optical and in the near-IR passbands defined by 
Bessel \& Brett (1988, AAO system).

The I-band shows the well-known quasi-constant absolute magnitude level for metallicities, 
[Fe/H], below -0.7. 
According to the models, the absolute J-band magnitudes are brighter than the I ones 
(an advantage) but also vary strongly, displaying a rapid monotonic increase with [Fe/H] 
(a disadvantage). 
The H-band magnitudes offer the advantage of being about 2 mag brighter, on average, 
than M$_I$. Unfortunately, 
the models indicate a complex dependence between [Fe/H] and the TRGB in H. 
M$_H$ exhibits a plateau at about -6, in the range -0.4$>$[Fe/H]$>$-1.3. Fig.~8 thus suggests that
an H-band TRGB might be useful at slightly higher metallicities than that for
which the I-Band TRGB is applicable. However, towards the
very low metallicities of interest for us, M$_H$ suddenly changes by
about 0.4~mag. Furthermore, the theoretical isochrones barely approach the
metallicity regime of our dwarf galaxy data. VII~Zw~403 in particular lies below
the lowest metallicity point covered by published models (see Fig.~11). 
(We add that we also investigated the TRGB magnitudes in our two grids, which
use the Bessell, Castelli \& Plez (1998) atmospheres rather than pure Kuruzc 
atmospheres adopted
in the Padova grids, and that they agree to within 0.05~mag).
  
Second, we compiled JHK data of GCs in the Milky Way and the
LMC from the literature. These data cover a large metallicity range. 
We used the observed CMDs to read off the magnitudes at the TRGB; 
together with the GC distances,
this yields another indication of the dependence of the absolute TRGB magnitudes
on metallicity. The major difficulty of this approach is that frequently, the
empirical RGBs of GCs are not sufficiently populated near
the tip to provide a reliable tip magnitude. 
Other sources of error include uncertainties in the GC distances and
metallicities. 

Kuchinski et al. (1995, a, b) observed several GCs
of the Milky Way belonging to the so-called disk
system. These have [Fe/H] metallicities between -1 and 0. The authors
also discuss older data for 47~Tuc and M~71. From these published CMDs,
we read off the K magnitudes of the TRGBs, and their H-K colors, to derive H-TRGB magnitudes.
We adopted the distance moduli and extinction corrections as presented
by Kuchinski et al. to derive absolute H-TRGB values for these 
nine clusters. Kuchinski et
al. use the Elias et al. (1982) standards for the CTI system. As they are
interested in disk clusters, extinction is a severe problem for some of the data
sets, especially for M~71 and Ter~2. The Kuchinski et al. cluster data have only a small
overlap in [Fe/H] with the dwarf galaxies we are interested in. 
But, they serve as a comparison with the models at high metallicity.

Ferraro et al. (1995) presented ground-based JHK photometry of stars in 12 
GCs of the LMC. The data are in the CTI system of Elias et al. (1982).
Unfortunately, the sampling at the TRGB is so poor in most of these
clusters, that a secure determination of the TRGB magnitude in either of the near-IR 
colors is not possible. Furthermore, only some of the 
GCs observed by Ferraro et al. have published [Fe/H] values,  
and these have rather large uncertainties.
We finally used only four of their GCs (see Fig.~11). The LMC clusters also have 
abundances which are higher than the range of interest for us, but, together 
with the Kuchinski et al. results, they allow an independent
check of the results derived from the isochrones. There is a rather large dispersion in the
observed data for these GCs (all with [Fe/H]$\ge$-1), 
but they do cluster about the
TRGBs of the isochrones, which is somewhat reassuring. 

Recently, Davidge \& Courteau (1999) observed four metal-poor Galactic halo GCs in the near-IR. 
They used the standard stars of Casali \& Hawarden (1992) and supplied values of the TRGB in
J, H, and K. The brighter parts of the RGBs are rather
well sampled.
The [Fe/H] values of these four clusters
range from -2.3 to -1.5. These are in the range of interest for studies of
metal-poor, old stellar populations in galaxies. Most importantly, they bracket
the abundance of VII Zw 403 ([Fe/H] = -1.92). The two more metal-rich GCs
of their sample overlap with the isochrone result, and there is good agreement between
the models and the data. These four GCs indicate
only a weak dependence of M$_{H, TRGB}$ on [Fe/H] for small values
of [Fe/H]. We use the four metal-poor GCs of Davidge \& Courteau, the model 
TRGB value for 15~Gyr and Z=0.0004, and our data point for VII~Zw~403, to derive M$_{H, TRGB}$. 
As can be seen from Fig.~11, a good approximation is 

\begin{center}
 $<$M$_{H, TRGB}$$>$~=~-5.5($\pm$0.1) \\
 for -2.3$<$[Fe/H]$<$-1.5.\\
\end{center}

We propose that this value may be used to derive the distances to dwarf galaxies 
containing metal-poor stellar populations, with
sufficient accuracy to be useful for stellar-population studies. 

We found disturbing the ``downward" trend of M$_{H, TRGB}$ for low [Fe/H] suggested
by the Padova isochrones, as compared with the ``flattening" of M$_{H, TRGB}$
in the range from -2.3$<$[Fe/H]$<$-1.5 indicated by the observations. We therefore
secured a stellar evolutionary grid at [Fe/H]~=~-2.3 from the Frascati group
(Cassisi, private communication). The 14~Gyr isochrone of their stellar-evolution
code yields M$_{H, TRGB}$~$\approx$~-5.6 at [Fe/H]~=~-2.3. This provides a consistency
check of our empirical result. We do not show this additional point in Fig.~8
and 11 since it is based
on a different stellar evolutionary code as the one to which we compare our data
throughout the remainder of the paper.

The TRGB is the preferred distance indicator when no observations of Cepheids
are available, because its physics is well understood, and because when
well calibrated (such as in the I band), it achieves a similar accuracy (Madore \& Freedman 1998.) 
The large number of stars populating the near-IR CMDs of star-forming dwarf galaxies near 
the TRGB (cf. also, Hopp et al. 2000) indicates that the TRGB method does not suffer from the 
statistical uncertainy which is sometimes encountered when investigating GC ridgelines 
in the near-IR. Therefore, the transformation and calibration uncertainties are the
dominant sources of error in using the near-IR TRGB as a distance indicator
for these galaxies. This method is superior to other 
distance indicators for galaxies in the 5-10~Mpc range; however, care must be taken not 
to mis-identify the tip-of-the-AGB (TAGB) with the TRGB.
The method of the three brightest RSGs, on the other hand, 
is always severely affected by small-number statistics
(Schulte-Ladbeck \& Hopp 1998, Greggio 1986). 

\subsection{M$_H$ as a measure of M$_{bol}$ for red stars} 

Bessell \& Wood (1984) investigated bolometric corrections for late-type stars.
From observations of individual red giants and supergiants in the MWG, LMC, 
and SMC, they find BC$_H$~=~2.6($\pm$0.2). In other words, the empirical value
for BC$_H$ is derived to be independent of color or spectral type, and metallicity.
In Figure~12, we show this empirical relation between M$_H$ and M$_{bol}$ as a straight 
line. An inspection of Fig.~5 in Bessell \& Wood (1984) suggests the possibility
that the lowest-metallicity data (those for the SMC stars) might be slightly
below their recommended value, at a BC$_H$ that is closer to 2.4, respectively.

To see how the recommendation of a constant BC$_H$ by
Bessell \& Wood compares to theoretical expectations, we overlay in Figure~12 
the locations of the TRGBs and TAGBs from
the Padova isochrone library. Their near-IR photometry is in the same
system as the data of Bessell \& Wood. We again used the 15~Gyr isochrones over the
range of available metallicities, and the values of M$_H$ and M$_{bol}$ associated
with these red giant models. These points lie near the empirical calibration,
but are clearly offset toward lower M$_H$ for a given M$_{bol}$. The lowest metallicity
point in the stellar-evolutionary models (Z=0.0004) corresponds to a BC$_H$ of 2.1. 
In order to make a comparison at high luminosities, we use the theoretical 
luminosities at the tip of the AGB. We show the TAGB points for solar metallicity 
and one fifth of solar, and for ages from 150~Myr to 15~Gyr. As illustrated in 
Fig.~12, while the stellar models with solar metallicity follow 
well enough the empirical relation, the lower-metallicity models are offset to a smaller
M$_H$ for a given M$_{bol}$.

The constancy of BC$_H$ in Bessell \& Wood is somewhat surprising,
since observed giants will be located
along the giant branches, and not just at the tips. In order to better
understand how BC$_H$ varies along the giant branches, we have
investigated in detail our transformed tracks. It turns out that in
the models, BC$_H$ increases along the tracks and that it depends on the
stellar mass (i.e. age of the population). These variations are
particularly pronounced for the Z=0.004 grid, with BC$_H$ at the tips
ranging between 2.3 and 2.6, and between similar limits 
for the 0.9~M$_\odot$ model in the brightest 1.5~mag of its RGB evolution.
The situation improves slightly for the Z=0.0004 grid, where BC$_H$ ranges
between just 2.0 and 2.2 at the various tips, and within approximately the same
limits for the 0.8~M$_\odot$ model in the brightest 1.5 mag of its RGB evolution.
A constant BC$_H$ is not justified from the models, and the interpretation of
transformations from the observational to the theoretical plane will require 
simulations, since there may be multiple choices for the theoretical parameters
of any individual star given its observed color and magnitude.

In summary, for the low metallicities which occur in our objects,
a smaller BC$_H$ than that suggested by the data of Bessell \& Wood appears to be 
more appropriate. While BC$_H$ clearly varies strongly with stellar mass, at Z=0.0004 the
TRGBs for models of a wide range of stellar masses can be approximated by 
a constant. We hence employ the following relation to characterize 
the dependence of M$_H$ on M$_{bol}$

\begin{center}
 $<$M$_{bol}$$>$~=~M$_{H}$~+~2.1($\pm$0.1) \\
 for Z~$=$0.0004\\
\end{center}

We use this relation to derive the bolometric luminosity function of the red stars in
VII~Zw~403 (Fig.~13.). This bolometric luminosity function is valid for the
stars at the TRGB (in terms of metallicity, color or temperature, BC$_H$), but is 
only a rough approximation (plus minus several thenths of a mag)
for other stars. We caution that using a constant BC$_H$ does introduce uncertainties
in going from the observed to the theoretical plane, and that the interpretation
of luminosity functions or HR diagrams (cf. also Alonso et al. 1999, Borissova et al. 1999)
requires simulations.

The TRGB in Fig.~13 is measured to occur at a bolometric luminosity of -3.4($\pm$0.1).
This agrees very well with the theoretically expected M$_{bol}$ from Fig.~8. 
Our result also compares very well with preliminary results from the DeNIS project by 
Cioni et al. (1999), who find M$_{bol, TRGB}$ to be -3.4 for the LMC and -3.6 for the SMC. 
These data were
transformed to M$_{bol}$ using different relations as H is not one of the survey bands.
The agreement is thus all the more encouraging. 

We detect a large number of stars below the TRGB.
Above the TRGB, the stellar numbers are small and thus highly affected by statistical errors. 
The bulk of stars which we observe in the luminosity range
 -3.4 $>$ M$_{bol}$ $>$ -6.5 is compatible with these stars being largely 
AGB stars in VII~Zw~403 (cf. also, Fig.~15). The data are consistent with the 
most luminous AGB stars being only a few hundred Myr of age, while the least luminous, 
oldest ones have ages of several Gyr.
The four most luminous red supergiants span the luminosity
range from approximately -6.5 $>$ M$_{bol}$ $>$ -8.5. We note that 
the cool part of the Humphreys-Davidson limit, the upper luminosity boundary of RSGs, is 
thought to occur at M$_{bol}$ of -9.5 (Humphreys \& Davidson 1994). The uncertainties
in the transformation from the observed to the theoretical plane prevent
us from stating how close our most luminous, near-IR detected objects
are to this boundary. A comparison of Fig.~14 with Fig.~5 reveals
that, while we are missing some of the bright supergiants in the near-IR as compared
to the optical, we do see the most luminous object, at M$_I$ of about -9, so we
are sampling the entire upper luminosity range. The reason why we miss in the near-IR
a few of the brightest supergiants, is due to the fact that the area of the NIC2 chip
covers less than a third of the star-forming centers, while the PC chip encompasses
these regions very well. 

The total absolute blue magnitude of VII~Zw~403 is small,
about -14 (see Schulte-Ladbeck \& Hopp 1998). At these small galaxy luminosities, fluctuations
in the numbers of the most luminous and hence most massive stars are expected (Greggio 1986).
Statistical effects have been known to be important in the UV and optical, 
and continue to be important for the interpretation of near-IR luminosity
functions and CMDs. We discourage the use of the brightest 
RSGs as a distance indicator. Similarly, the comparative studies of the RSG
populations of galaxies are severely affected by small-number
statistics (cf. the stellar frequencies for M$_{bol}$ $<$~-7 in Fig.~17 of Massey 1998).

\subsection{Decoding the near-IR CMDs} 

In Fig.~14, we color-code the main areas of the CMDs using terminology
that an observer would employ to classify stars according to the different
morphological features seen on the CMD.
The classification is based on the [(V-I)$_o$, M$_{Io}$] CMD of Fig.~5.
The stars marked in blue are mainly MS and BSG stars; 
stellar evolutionary tracks tell us BL stars occur here as well. 
The stars indicated in magenta are considered to be RSG; the 
tracks suggest there may be BL stars at the faint end. More importantly,
the tracks indicate bright AGB stars populate this part of the red
plume as well, and we cannot easily differentiate them from RSGs. 
We are able to clearly distinguish AGB stars
when they form the red tail; these stars are marked in black. Finally,
we colored all of the data points which occur below the TRGB in
red, suggesting that mainly RGB stars are found here. However, stellar
evolution tracks indicate we must be aware of BL and faint AGB stars
in this part of the CMD as well. With this broad classification of
stars, we now compare the morphology of the optical--near-IR CMDs.

In Fig.~14, we use the stellar classification to investigate the
stellar content of the [(J-H)$_o$, J$_o$] and [(J-H)$_o$, H$_o$] CMDs.
This elucidates a problem which we alluded to earlier, namely that even those 
AGB stars which populate the redward-extended tail of optical CMDs
overlap with the RSGs in near-IR CMDs. As we stated before, the colors
of RSG, AGB and RGB stars are quite degenerate with respect to (J-H)$_o$
color, or temperature. Folded with the color errors that arise from the
measurements, only luminosity can help us distinguish
between RSG and bright AGB stars on the one hand, and RGB
(plus faint AGB stars) on the other hand. At the top of the red plume,
we can distinguish the most luminous RSGs from the bright AGBs based
on luminosity (AGB stars are not expected at an M$_{bol}$ of -8, cf. Fig.~12).
We notice from Fig.~14 some mingling of faint stars between blue
and red colors. This is most pronounced for stars below the
TRGB. Above the TRGB, only a few stars are seriously ``misplaced" based
on optical color in the near-IR CMDs. We also note that our placement of
the TRGB derived from the luminosity functions in F110W and F160W is consistent
with our stellar classification scheme. In other words, no stars classed
RGB stars based on the [(V-I)$_o$, M$_{Io}$] migrated above the TRGB
limit in the near-IR CMDs, while just a few stars classed RSG or AGB stars
wander below the TRGB. Hence, there is an overall good agreement
between the TRGB magnitudes based on stellar classification and the luminosity
functions.

We may ask what kind of stars are we missing in the optical
CMDs which are present in near-IR CMDs. Comparing Fig.~14 with Fig.~10,
some differences for faint stars are apparent, in the sense that some faint
stars of extreme color in Fig.~10 do not appear in Fig.~14. This is not
unexpected as objects that are extremely red may have been missed in the
optical, and objects that are extremely blue may be blends or spurious detections
in the near-IR. There is very good agreement for the brightest, and hence most
luminous objects -- it appears that none of the brightest supergiants seen
in J and H were not seen in V and I as well, giving some confidence that
we have a complete sample of the brightest supergiants that were encompassed
by the NIC2 chip's area in the near-IR.
The only potentially significant difference then
between Fig.~10 and 14 occurs in the color range (J-H)$_o$$>$~1.1.
Five stars are found here with very red colors, and with brightnesses above
the TRGB. They have small measurement errors and must be considered to be
real. In this parameter space, we thus picked up additional objects in the
near-IR photometry. It is possible, judging from their colors, that these 
objects are Miras or Carbon stars.

In Fig.~15, we show a CMD that uses the I and H bands. It may be compared with
Fig.~5 which employs the I and V bands. Fig.~15 illustrates that our expectations for observing
in the near-IR were realized: the H-band allowed for
an over 1~mag gain in stellar brightnesses for the red stars. The tracks cross the
data in slightly different places on the [(V-I)$_o$, M$_{Io}$] and the [(I-H)$_o$, M$_{Ho}$] 
planes, with offsets of a few tenths of a mag. These discrepancies are not too surprising 
considering the uncertainties
in the transformations from the space to the ground systems on the one hand,
and the difficulties of model atmospheres to reproduce empirical color-temperature
relations on the other hand. They suggest that the state-of-the-art for comparing
empirical and synthetic CMDs is to reproduce the gross features of the stellar
distibutions including the absolute magnitude of the TRGB quite well, whereas
other quantitative results 
such as e.g., metallicities from the location of the RGB, or detailed SFHs, should 
be considered somewhat more uncertain. 

\subsection{Long color-baseline CMDs} 

We now present and interpret CMDs with long color baselines. 
Fig.~16 shows the CMDs of I$_o$ or M$_{Io}$ versus (V-I)$_o$,
(V-J)$_o$, and (V-H)$_o$. The same color scheme as that used
for Fig.~14 is employed; however, notice the change in color scale.
These CMDs in principle offer advantages for
separating out different stellar phases from one another. In practice,
since the photometric errors vary from 
band to band and depend on the
colors of the objects, the potential of these CMDs to distinguish 
different stellar phases is not fully realized.

An impressive feature of the CMDs of Fig.~16 is the increasingly
larger spread of the data in color. For instance, the data points
in the (J-H)$_o$ CMD of Fig.~10 subtend only about 1.5~mag, those
in the (V-I)$_o$ CMD already cover 4.5~mag. In (V-J)$_o$ this baseline has
grown to about 6.5~mag, and in (V-H)$_o$, the data are distributed over a range
of almost 8~mag. 

As we study CMDs with increasing color baseline, the color errors at the
faint end of the stellar distributions become very large. Therefore, for the
faint magnitudes, we observe some mingling of stars across the CMDs. 
We do not consider this to be a real effect. Only one bright star  
migrates from the red plume of the (V-I)$_o$ CMD into the blue
plume of the (V-H)$_o$ CMD. This could potentially be a heavily reddened
blue object, or else a blend of unresolved objects weighted very differently
in different colors.

The red plume of the (V-I)$_o$ CMD was considered to be composed of RSGs plus AGBs
mainly in a narrow region where it appears to form a linear sequence from low to
high luminosity. Stars offset to red colors from this band were classed AGBs only. 
Comparing the three CMDs with each other, we see that we judged quite
well from the (V-I)$_o$ CMD the location of the AGB. Two luminous objects
considered RSGs based on the (V-I)$_o$ CMD might also be classed AGB stars 
based on the (V-H)$_o$ CMD. At least half a dozen or so of the faint red stars
near the dividing line between the RSG, AGB, and RGB stars in the
(V-I)$_o$ CMD could be additional AGB stars based on the (V-H)$_o$ CMD.
In Fig.~14, two BSGs wandered from the blue plume into the red plume
of the CMDs, demonstrating further difficulties to disentangle the
nature of red stars in this luminosity range based on CMDs.

A feature of the four brightest RSGs detected in all four bands is that they
appear to develop a progressively larger redward color offsets in the (V-J)$_o$ 
and (V-H)$_o$ CMDs from the linear band that was used to class them as RSGs
on the (V-I)$_o$ CMD. Since these are bright objects with small measurement
errors, we assume that this effect is real. We do not expect to see AGB stars
at such high luminosities, and so must assume that these colors are intrinsic
to the most luminous RSGs sampled by the NIC2 frame.

A significant spreading out of colors occurs for the bluest stars and stars between
the MS and the RSG plume/red tangle; presumably high- and intermediate-mass 
stars on blue loops. This is in part due to high measurement errors
for blue objects in the near-IR bands. The effect
of differential reddening could play a role in this 
re-distribution of sources, but we cannot constrain it on the basis of
our data. 

\subsection{Comments on the Blue Hertzsprung Gap and the blue-to-red
supergiant ratio at low metallicity}

An interesting feature of the optical/near-IR CMDs of Figs.~14-16 is the
appearance of a gap in the distribution of stars along the blue plume.
We first noticed this in the optical, PC data centered on the star-forming
regions of VII~Zw~403 (SCH98). We interpret this gap as the Blue Hertzsprung Gap (BHG) 
which is predicted by stellar-evolution theory, to occur between the distribution 
of massive stars at the
red edge of the MS (core-H burning) and the blue edge of the BL phase (core-He burning). 

Simulations of the VII~Zw~403 CMD with the code of Greggio et al. (1998), which use the
low-metallicity grid also employed in this paper,
were presented in SHCG99. The synthetic CMDs shown as Fig.~6 in SHCG99 display the 
predicted BHG for young ages of the stellar population. In comparing Fig.~6
of SHCG99 with the CMDs presented in this paper, it appears that the BHG
occurs at a lower luminosity in the models than that feature which we
identify with the BHG in the data. This difference can readily be explained
recalling that the simulations were carried out for stellar evolution at
Z=0.0004 (Z$_\odot$/50). The extent of the BLs in stellar evolution
models is very sensitive to metallicity. In Fig.~15, we connect with a straight
line the locations of the blue edges of the blue loops for the 20 and the 9~M$_\odot$
tracks in the Z=0.0004 and the Z=0.004 (Z$_\odot$/5) grids. These nicely bracket the
location of the observed BHG. The appearance of the BHG between the tracks of these two 
grids is consistent with the metallicity of the ionized gas of VII~Zw~403,
which suggests the present generation of stars has metallicities closer
to Z$_\odot$/20. In future simulations, we will incorporate into the
code the Z=0.001 metallicity grid, as well as all the bands now available
from observations. 

The BHG is seen very well in the CMD of the nearby Local Group dIrr Sextans~A
(Dohm-Palmer et al. 1998 and references therein), for which the WFPC2 photometric
errors are even smaller in the [(V-I)$_o$, M$_{Io}$] CMD than those for the more
distant VII~Zw~403. The oxygen abundances
for Sextans~A (log~(O/H) = -4.48, Skillman 1989) and VII~Zw~403 (log~(O/H) = -4.42(0.06), 
Martin 1997, and -4.31(0.01), Izotov, Thuan \&  Lipovetsky 1997) are quite similar,
around Z$_\odot$/20. A gap in the distribution of stars was {\it not seen} in the H-R
diagram of the slightly metal-poor (Z$_\odot$/3) LMC by Fitzpatrick \& Garmany (1990). 
Since it was predicted
by stellar-evolution theory, its absence in these data gave rise to the problem
of the {\it missing BHG} (e.g. the reviews of Maeder \& Conti 1994 
and Chiosi 1998). Theoretical work has therefore been aimed at filling in the ``missing" BHG
with stars, for instance by developing and incorporating into the theory of stellar evolution
new prescriptions for internal mixing and/or mass loss in massive stars (Salasnich, Bressan,
\& Chiosi 1999). We note that the BHG is now also being discovered in
CMDs of the LMC (Zaritsky 1999, private communication) which are being assembled as part
of the digital Magellanic Could Photometric Survey (Zaritsky, Harris \& Thompson 1997).
These data will give insight into why the BHG was not observed by Fitzpatrick \& Garmany (1990);
first suspicions include the smaller sample size and the spatial distribution of stars
across the LMC.

The fact that {\it the BHG has now been seen in the CMDs of several metal-poor
galaxies} indicates that more observational work on the HR diagrams of
metal-poor star-forming galaxies is needed, and that such CMDs may provide important 
guidance for stellar-evolution theory. They suggest that additional mixing
and mass-loss may not be needed to the extent currently anticipated by theorists.

A related issue is the blue-to-red supergiant ratio in galaxies. Recent
reviews/papers on this open problem of massive-star evolution are those of 
Langer \& Maeder (1995) and Deng, Bressan \& Chiosi (1996). 
In brief, the B/R supergiant ratio as a function of
Z has not yet been predicted consistently by stellar-evolution models. However, the data
being used to address this problem largely date back quite some time now, to the 
excellent series of papers by Humphreys et al. on the supergiant stellar content
of Local Group galaxies (e.g., Humphreys \& McElroy 1984). 
Our data allow us to contribute a new measurement at low-metallicity, but only in
a limited way; again, small-number statistics rules. 
Since the largest problems for predicting the B/R ratios generally occur at low
metallicities, the limited insight gained here may nevertheless be useful.

Owing to the visibility
of the BHG on the CMDs, we are quite well able to separate the core-H burning MS
from the core-He burning BSG stars. Counting stars in the various CMDs shown in
this paper yields a number of about 20$\pm$3 BSGs, with an RMS error of 4.
We anticipate that the largest systematic error arises from preferrentially missing
BSGs due to crowding in the 
star-forming centers; this would have the effect of increasing the B/R ratio.
Assessing the number of (core-He burning) RSGs turns out to be more difficult.
This is owing to the already much deliberated fact that it is difficult to
discern massive RSGs from intermediate-mass, AGB stars. This is true
for our photometry as well as for the other existing data sets which
attempt to address the B/R ratio (Brunish, Gallagher \& Truran 1986).
Based on their high luminosity we can clearly identify 4 stars with RSGs. 
In this case, assuming
the counts are dominated by the RMS errors in the stellar numbers,
the B/R ratio for VII~Zw~403 is 5$\pm$4. This ratio is consistent with the SMC (Z$_\odot$/10)
numbers (cf. Langer \& Maeder 1995). If additional stars in the red plume of 
VII~Zw~403 are true red supergiants in the evolutionary sense, then the 
ratio goes down. This demonstrates the large effects that statistical
and systematic errors have in affecting empirical assessments of the B/R supergiant ratio.

\subsection{From CMDs to integrated photometry --- implications for
detecting old stellar populations}

In this section, we perform an exercise aimed at clarifying
the interpretation of integrated photometry of BCDs. For this purpose, 
consider the NIC2 chip to be a single aperture photometer of about 19" x 19" 
centered on the active star-formation regions of the BCD. 
As discussed in the previous section, HST single-star photometry 
and stellar-evolution tracks allow us to distinguish broadly the main different
stellar phases on the CMDs of VII~Zw~403. We add the light of stars
in different phases to investigate their contributions
to the total light in the V$_o$, I$_o$, J$_o$, and H$_o$ bands. In other words, we perform
population synthesis based on single-star colors and luminosities.

The results of the summation are shown in Table~2. We divide the stars into
three categories: 
stars in the blue plume (MS, BSG and BL stars), red stars above the TRGB 
(RSG and bright AGB stars), and red stars
below the TRGB (RGB stars, some faint AGB stars, and possibly some BL stars). 
Among the stars 
below the TRGB, the dominant population is considered to be RGB stars. 
However, depending on the age and early SFH of this BCD (which we cannot 
completely determine  
from the morphology of the red tangle alone) there may also be  
faint AGB stars. Because we are positioned on the star-forming centers, some
faint BL stars might contribute here; for the less massive stars, the blue
and red portions of their evolution are not as widely separated on our diagrams
as those for the more massive stars.

Looking at Table 2, one simple result is clear: 
{\it A few luminous stars outshine even a large number of faint stars.} 
The light in an aperture centered on the star-forming regions
is dominated in the I$_o$, J$_o$, and H$_o$ bands by RSG and luminous AGB stars. 
These obviously outshine the older RGB stars. The MS, BSG and 
BL stars belonging to the young and blue population
are also significantly brighter in all three
near-IR bands than the old stellar population. The sea of red giants
so well resolved with HST contributes less than 15\% to the integrated
light in any of these filters. 

Notice also that 
three quarters of the light in V$_o$ comes from the young, blue stars, 
and less than 10\% from the evolved stars, reinforcing the results of
Schmidt, Alloin \& Bica (1995) that even a small mass fraction
of young stars superimposed on the background sheet of
a dwarf Elliptical (dE) or dSph galaxy will completely outshine and render undetectable
the underlying old population in optical passbands. 

Our results have implications for the integrated photometry
of BCDs in general. 
The actively star-forming regions of BCDs are the visually brightest regions. 
Aperture photometry, including that in the near-IR 
(e.g. Tully et al. 1981, Thuan 1983, 1985), was traditionally 
carried out with photometric 
apertures centered on these bright H-II regions. The apertures used were usually
about 10" in size, thus encompassing mainly the star-forming centers.
As our exercise suggests, such observations are dominated by the 
young (less than about 50 Myr old) red supergiants 
from the most recent star formation activity and 
by luminous AGB stars with ages less than a few Gyr. 
We concur with James (1994), 
that luminous AGB stars must be considered an important
component of BCDs, contributing around half of the light in the star-forming region,
and reflecting star-forming activity in the last few Gyr. 

It is therefore not possible to conclude, from integrated photometry centered on 
the starburst alone, whether an underlying, old ($>$10~Gyr) stellar population is present. 
Since even the near-IR bands 
predominantly measure the light from the younger stars, 
forming colors such as V-J or V-H which involve long baselines 
does not help to break this ambiguity.

In Schulte-Ladbeck et al. (2000), we address the possibility to use
spectrophotometric indices based on long-slit spectra that exclude
the star-forming regions (see Hopp, Schulte-Ladbeck \& Crone 1998) 
in order to age-date the background sheets
of BCDs. This method has its roots in the work on spectrophotometric
dating of Elliptical galaxies (e.g. Worthey 1994); and while it suffers
from its own ambiguities, it may represent an important alternative to single-star
photometry for dating distant BCDs in the future.

The morphological properties of VII~Zw~403 are representative of the
vast majority of BCDs. We have previously demonstrated
(SHCG99) the presence of a ``core-halo" morphology for the resolved stars in
this galaxy. Young stars
are exclusively found in the core near the center of the extended
light-distribution of the halo (approximately in the inner
40"). At large radii (out to 100" or about 4 disk scale lenghts), 
this young population is absent, but we find some AGB stars
and a very strong red tangle dominated by RGB stars.
As discussed in Schulte-Ladbeck \& Hopp (1998),
stellar population synthesis of integrated colors in the outer halos of BCDs 
(using color gradients derived from multi-filter surface photometry) suggest
that even here, a mixed-age population (resulting from
a complex SFH) is consistent with the data. 
In the case of VII~Zw~403 where we
know from single-star photometry that bright RSGs 
are absent in the halo, we have interpreted the halo color profile
as evidence for the presence of an underlying population which is at least a few Gyr old
(and possibly truly ancient, $>$~10~Gyr old). 
It is likely that the halos of other iE BCDs are similarly composed of 
old and/or intermediate-age stars. 

One of the most interesting implications of our results here; however,
is that those BCDs which do {\it not} exhibit an outer halo,
do not necessarily lack old stars.  It is entirely possible that their star forming regions
are scattered through an older populations, totally obscuring it.

\section{Summary and Conclusions}

The quality of the NICMOS photometry in terms of crowding and limiting magnitudes 
represents an improvement by several apparent magnitudes over exisiting ground-based
near-IR data of similar galaxies. In terms of absolute magnitudes, comparable data are only 
available for the MCs. The results of our single star photometry demonstrate that all 
of the important stellar phases of composite stellar populations can be traced 
on near-IR CMDs. The optical/near-IR CMDs of VII~Zw~403 show the ``missing"
Blue Hertzsprung Gap and a blue-to-red supergiant ratio of about 5, 
providing new input for stellar-evolution models at low metallicity.

Several steps could be undertaken to improve our analysis:
1)~Modeling of stellar atmospheres and stellar evolution at the low metallicities appropriate
for the earliest stellar populations in dwarf galaxies, 2)~Additional modeling and
observations of the AGB stellar phase, to better understand the effects of 
AGB ages and 
metallicities, 3)~Near-IR observations providing a uniform set of cluster 
CMDs in a well-established JHK system, for comparison of such simple stellar
populations with stellar theory and galaxy observations.

Our data reach the RGB of VII~Zw~403 to completeness levels that allow 
a secure measurement of the J, H magnitudes at the tip. We compare
our results to that of clusters, DeNIS data of the MCs, and the Padova tracks
of stellar evolution, and give a conversion from apparent to absolute
TRGB magnitude. Providing a TRGB fiducial in the near-IR
has advantages for estimating the distances to more distant galaxies. 
As red giants are observed near the peak of their energy distributions, the near-IR
TRGB method can potentially reach to further distances than
the optical TRGB method. It may also prove useful for  
highly reddened galaxies. 

The bolometric luminosity function of the red stars shows that for any
stellar component other than the red giants, the RMS error in the star counts is
large, even in the near-IR. Statistical fluctuations in stellar numbers 
in low-luminosity galaxies are well known to become large for massive stars. 
We caution against using the method of the brightest red supergiants as a distance
indicator even in the near-IR.

Summing up the light of the resolved stars in different evolutionary phases, 
we illustrate that published integrated photometry of BCDs, 
in both the optical and the near-IR, is dominated by the light from luminous, 
young red supergiants from the current starburst (the last
50~Myr or so) and luminous young and intermediate-age AGB stars. 
Attempts to distinguish an old stellar component using  near-IR colors 
are bound to fail because of the 
color degeneracy among RSG, AGB, and RGB stars. 
Even a large number of red giants is 
undetectable underneath just a few luminous RSG and/or AGB
stars. This shows why 
contradicting conclusions have been reached in the 
literature regarding the oldest stellar components of BCDs. 

The purpose of our study is to use SFHs to answer very general questions about
the nature of BCDs:  Are they ``young" galaxies?  Are they related to the
faint blue excess? 
We note that AGB stars are clearly identified in the 
resolved stellar content, and also contribute significantly to the integrated near-IR light of
VII~Zw~403. The presence of an intermediate-age stellar population suggests that
star formation was active in this galaxy at times which correspond to reshifts of a few tenths.
VII~Zw~403 is clearly not a young galaxy. 
We argue that the AGB stars may provide a further link between the type 
iE BCDs and the faint-blue galaxy population. The trick will be to determine whether 
or not this component of AGB stars can correctly account for the right amount of star 
formation at the right time, to identify iE galaxies as the (non-merged) remnants of
the (CNELG) faint blue excess.  

In conclusion, we present the first exploration of the near-IR CMD of a BCD 
galaxy, to deep enough limiting magnitudes to detect the
evolved descendents of low-mass stars. Together with recent results regarding
the structural parameters of BCDs (e.g. Sung et al. 1998), our data provide further support 
of the idea that type 
iE BCDs possess dynamically relaxed, old stellar populations. The nature of the iE BCDs 
is inferred to be one of flashing dEs/dSphs. An interrelated evolution between 
BCDs and dEs/dSphs galaxy types (via gas cycling through starbursts) could potentially 
account for precursors as well as remnants of the faint blue galaxies.

\acknowledgments We acknowledge financial support through 
archival HST research grants to RSL (AR-06404.01-95A, AR-08012.01-96A) and guest 
observer HST grants to RSL and MMC (GO-7859.01-96A). This project 
benefitted from the hospitality of the Universit\"{a}tssternwarte 
M\"{u}nchen, where RSL found a stimulating environment during
her sabbatical leave from the University of Pittsburgh. 
We are grateful to Dr. M. Dickinson for sharing his time and software 
to help us remove the pedestal in the NICMOS data. We also thank 
D. Gilmore for excellent support during our visit to STScI. 
We further thank Dr. L. Origila for sharing her HST magnitudes of the 
Bessell, Castelli \& Plez (1998) stellar atmospheres with us prior to publication. 
UH acknowledges financial support from SFB375.

\clearpage
\noindent Figure Captions

\figcaption[]{The NIC2 images after pedestal removal and re-calibration
are displayed on the left, with F110W on top and F160W at the bottom.
The horseshoe-shaped H-II region and the highly inclined background spiral
help to identify the area covered by the NIC2 images w.r.t. the WFPC2
images, cf., Plate~1 of SCH98 for our color image of the UV/optical PC data. 
The images on the right illustrate the appearance of the images after 
background-smoothing was applied; a smooth background is essential 
for DAOPHOT single-star photometry.}

\figcaption[]{Zooming in on the images in the F110W (top) and F160W 
(bottom) filters. On the left, the original PSF of the images is
retained, on the right, the images on which we did the photometry, 
with PSF smoothing applied.}

\figcaption[]{The photometric errors computed by DAOPHOT. }

\figcaption[]{The fraction of test stars recovered on the
NIC2 chip. }

\figcaption[]{Color-magnitude diagram of the WFPC2 observations in V and I,
corrected for foreground extinction. Overplotted are selected Padova tracks
for a metallicity of Z=0.0004 and Z=0.004. The salient morphological features of this
CMD are the blue plume, which contains MS, BSG, and BL stars, and the
red plume, which at high luminosities, is comprised of RSG and AGB stars,
and at low luminosities, becomes the well populated red tangle which
contains RGB, low-luminosity BL, and low-luminosity AGB stars. 
The extended red tail is comprised only of AGB stars.}

\figcaption[]{Color-magnitude diagrams of near-IR magnitudes
in the Vega system. All stars found in F110W and F160W (998) are plotted.
The near-IR CMDs exhibit a blue plume, as well as  a red plume with a densely
populated red tangle at low luminosities. An extended red tail is not seen.}

\figcaption[]{Two-color diagrams for various combinations of optical
and near-IR photometry. The points shown have photometric (1~$\sigma$) errors smaller
than 0.15~mag.}

\figcaption[]{The absolute magnitudes at the TRGB were read off from
the 15~Gyr isochrones of the Bertelli et al. (1994) stellar models
for metallicities of (from left to right), Z=0.0004, 0.001, 0.004,
0.008, 0.02, 0.05. The color-coding from top to bottom gives the tips
in H (black), J (red), I (magenta), and bolometric (blue) magnitudes.}

\figcaption[]{The luminosity functions in F110W (top) and F160W
(bottom) for stars with 0.75$<$(F110W-F160W)$<$1.5. The location of the
TRGB is marked. }

\figcaption[]{Color-magnitude diagrams of near-IR magnitudes
using the transformations described in
the text. The bottom panel has overplotted the same tracks as those used
in Fig.~5.}

\figcaption[] {The absolute magnitude at the TRGB derived for
VII~Zw~403 (in red) is compared with those of stellar models (in green)
and of globular clusters (in black). The open circles are clusters
from Ferraro et al., no error bars are indicated owing to the difficulty
of reading off the TRGB. The filled circles above [Fe/H] of -1 are from Kulchinski et
al.; here, it was possible to estimate an error bar. The error bars reflect
how well the RGB is populated near the tip, and how well, as a consequence, we
felt that we could determine the location of the TRGB from these data.
The points at low metallicities are the cluster data of Davidge \& Courteau;
we adopted their errors. }

\figcaption[] {The relationship between M$_H$ and M$_{bol}$ for red stars. The straight lines
come from Bessel \& Wood (1984) and represents their fit to data and associated error bars
of observations of a variety of red giants and supergiants in the Galaxy and the MCs.
In green, we display the TRGBs of the Bertelli et al. (1994) 15~Gry isochrones
for the full range of metallicities. The lowest metallicity point is the one
with the smallest M$_H$; and it suggest a bolometric correction of 2.1. 
To extend the luminosity range, we also 
read off the tips of the AGBs on model isochrones.
We did this for isochrones of two different metallicities, solar and one fifths of
solar, and for ages ranging from about 150~Myr to 15~Gyr. An extensive discussion of
the bolometric corrections derived from individual stellar models
of different masses is provided in the text.}

\figcaption[] {The absolute, bolometric luminosity function, as derived
from the H-band luminosity function and using a constant to transform from M$_H$
to M$_{bol}$ (see the text). Notice that the use of a constant bolometric
correction introduces substantial (of order of several tenths of a mag) uncertainties.
A unique mapping of observational data onto the theoretical plane may not be
possible, and solutions need to be investigated with simulations. The feature that
can be interpreted fairly securely is the TRGB at M$_{bol}$~=~-3.4, as we set up
our transformation to be appropriate for these stars.}

\figcaption[]{Application of the stellar classification scheme based on the optical
CMD to the near-IR CMDs. We use color-coding to
illustrate the location of specific stellar-evolutionary phases. The color-coding
is discussed in the text, and the labels given in the top CMD reflect
what we judged to be the main stellar contributor to each morphological feature 
based on this CMD. The location of the TRGB is marked by a dashed
green line. Only stars detected in four colors are shown. 
Although we observe a limited number of stars to cross
the TRGB line in luminosity, in general, there is good agreement between 
the TRGB location as derived based on the luminosity functions and the location
of the TRGB as indicated by our stellar classification scheme. We can also see that
the RSG and AGB stars overlap closely in color and to a large degree, in
luminosity as well, on near-IR CMDs.}

\figcaption[]{Color-magnitude diagram of near-IR colors with the same tracks as
those used in Fig.~5 overplotted. An absolute-magnitude scale in terms of M$_H$
is provided as well. Comparing this CMD with that of Fig.~5 illustrates 
how well our expectations were realized -- we gained a little over 1~mag by observing 
the red stars in near-IR. We marked by thick bars the connecting points of the
blue edges of the BL phases for the 20 (left) and the 9 (right) M$_\odot$ stellar models 
in the two
metallicity grids. They bracket the gap in the blue plume also seen in Figs.~13 and 15.
This gap is interpreted to be the ``missing" Blue Hertzsprung Gap.}

\figcaption[]{Color-magnitude diagrams of stars
found in all four bands. The I magnitude provided the original TRGB distance scale
and the TRGB shown by the dashed green line is on the original absolute magnitude scale.
The progressively longer color baselines illustrate how different stellar phases can
in principle be well separated using near-IR colors, even for early-type stars. In practice,
the errors for stars of different color vary sufficiently across the three CMDs to counteract
some of these advantages in the present data set.}


\begin{references}


\reference{ } Alonso, M.V., Minniti, D., Zijlstra, A.A., Tolstoy, E.
1999, A\&A, 346, 33
\reference{ } Allen, D.A., Cragg, T.A. 1983, MNRAS, 203, 777
\reference{ } Babul, A., Ferguson, H.C. 1996, ApJ, 458, 100
\reference{ } Bedding, T.R., Minniti, D., Courbin, F., Sams, B. 1997,
ApJ, 495, L13
\reference{ } Bertelli, G., Bressan, A., Chiosi, C., Fagotto, F., Nasi, F. 
1994 A\&AS, 106, 275 
\reference{ } Bessell, M.S., Brett, J.M., 1988 PASP, 100, 1134
\reference{ } Bessell, M.S., Castelli, F., Plez, B. 1998, A\&A, 333, 231
\reference{ } Bessell, M.S., Wood, P.R., Brett, J.M., Scholz, M. 1991,
A\&AS, 89 335
\reference{ } Bessell, M.S., Wood, P.R. 1984, PASP, 96, 247
\reference{ } Borissova, J., Georgiev, L., Rosado, M., Kurtev, R., Bullejos, A.,
Valdez-Guiterrez, M. 1999, A\&A preprint (astro-ph/9903019)
\reference{ } Brandl, B., Sams, B.J., Bertoldi, F., Eckart, A., Genzel, R.,
Drapatz, S., Hofmann, R., Loewe, M., Quirrenbach, A. 1996, ApJ, 466, 254
\reference{ } Brunish, W.M., Gallagher, J.S., Truran, J.W. 1986, AJ, 91, 598
\reference{} Campbell, A.W., Terlevich, R. 1984, MNRAS, 211, 15
\reference{} Cardelli, J.A., Clayton, G.C., Mathis, J.S. 1989, ApJ, 345, 245
\reference{} Casali, M., Hawarden, T. 1992, JCMT-UKIRT Newsl., No. 4, 33
\reference{} Chiosi, C. 1998, in ``Stellar Astrophysics for the Local Group",
eds. A. Aparicio, A. Herrero, and F. S\'{a}nchez (Cambridge University Press), p. 1
\reference{} Cioni, M.R., Habing, H.J., Loup, C., Groenewegen, M.A.T., Epchtein, N., and the
DeNIS Consortium 1999, in IAU Symp. 192, ``The Stellar Content of Local Group
Galaxies", eds. P. Whitelock, R. Cannon, (PASP: IAU) p. 65
\reference{ } Davies, R.I., Sugai, H., Ward, M.J. 1998, MNRAS, 295, 43
\reference{ } Davidge, T. J., Courteau, S. 1999, AJ, 117, 1297
\reference{ } Deng, L., Bressan, A., Chiosi, C. 1996, A\&A, 313, 159
\reference{ } Dohm-Palmer, R.C., Skillman, E.d., Saha, A., Tolstoy, E., Mateo, M., Gallagher, J.,
Hoessel, J., Chiosi, C., Dufour, R.J. 1998, AJ, 115, 152
\reference{ } Dekel, A., Silk, J. 1986, ApJ, 303, 39
\reference{ } Ellis, R.S. 1997, ARA\&A, 35, 389
\reference{ } Elias, J.H., Frogel, J.A., Matthews, K., Neugebauer, G. 1982, AJ, 87, 1029
\reference{ } Elias, J. H., Frogel, J. A., Hyland, A. R., Jones, T. J. 1983, AJ, 88, 1027
\reference{ } Fagotto, F., Bressan, A., Bertelli, G., Chiosi, C.
1994, A\&AS, 104, 365
\reference{ } Ferguson, H.C., Babul, A. 1998, MNRAS, 296, 585
\reference{ } Ferraro, F.R., Fusi Pecci, F., Testa, V., Greggio, L., Corsi, C.E., 
Buonanno, R., Terndrup, D.M., Zinnecker, H. 1995, MNRAS 272, 391
\reference{ } Fitzpatrick, E.L., Garmany, C.D. 1990, ApJ, 363, 119
\reference{ } Greggio, L. 1986, A\&A, 160, 111
\reference{ } Greggio, L., Tosi, M., Clampin, M., De Marchi, G.,
Leitherer, C., Nota, A., Sirianni, M. 1998, ApJ, 504, 725
\reference{ } Guzm\'{a}n, R., Jangren, A., Koo, D.S., Bershady, M.A.,
Simard, L. 1998, ApJL, 495, 13
\reference{ } Holtzman, J.A., Burrows, C.J., Casertano, S., Hester, J.J., Trauger, J.T.,
Watson, A.M., Worthey, G. 1995, PASP, 107, 1065
\reference{ } Hopp, U., Schulte-Ladbeck, R.E. 1995, A\&AS, 111, 527
\reference{ } Hopp, U., Schulte-Ladbeck, R.E., Crone, M.M. 1998,
in Bonn/Bochum-Graduiertenkolleg Workshop, ed. T. Richtler \& J.M.
Braun (Shaker Verlag), 161
\reference{ } Hopp, U., Schulte-Ladbeck, R.E., Greggio, L., Crone, M.M. 2000, Proc. STIS
Workshop on ``Spectrophotometric Dating of Stars and Galaxies", eds. I. Hubeny, S. Heap, and
R. Cornett (PASP), in press
\reference{ } Humphreys, R.M., Davidson, K. 1994, PASP, 106, 1025
\reference{ } Humphreys, R.M., McElroy, D.B. 1984, ApJ, 284, 565
\reference{ } Izotov, Y.I., Thuan, T.X., 1999, ApJ, 511, 639
\reference{ } Izotov, Y.I., Thuan, T.X., Lipovetsky, V. 1997, ApJS, 108, 1
\reference{ } James, P.A. 1994, MNRAS, 269, 176
\reference{ } Johnson, H.L. 1966, ARA\&A, 4, 193
\reference{ } Koornneef, J. 1983, A\&A, 128, 84
\reference{ } Kuchinski  L.E., Frogel, J.A., Terndrup, D.M., Persson, S. E. 
1995, AJ, 109, 1131
\reference{ } Kuchinski, L.E., Frogel, J.A. 1995, AJ, 110, 2844
\reference{ } Kunth, D., Martin, J.M., Maurogordato, S., Vigroux, L. 1986,
in ``Star-Forming Dwarf Galaxies and Related
Objects", eds. D. Kunth, T.X. Thuan, and J. Tran Than Van 
(Gif-sur-Yvette: \'{E}ditions Fronti\`{e}res), 89
\reference{} Kunth, D., S\`{e}vre, F., 1986, in ``Star-Forming Dwarf Galaxies and
Related Objects", eds. D. Kunth, T.X. Thuan, and J. Tran Than Van 
(Gif-sur-Yvette: \'{E}ditions Fronti\`{e}res), 331
\reference{ } Langer, N., Maeder, A. 1995, A\&A, 295, 685 
\reference{ } Lee, M.G., Freedman, W.L., Madore, B.F., 1993 ApJ, 417, 553 
\reference{ } Loose, H.H., Thuan, T.X. 1986, in ``Star-Forming Dwarf Galaxies
and Related Objects", eds. D. Kunth, T.X. Thuan, and J. Tran Than Van
(Gif-sur-Yvette: \'{E}ditions Fronti\`{e}res), 73
\reference{ } Lynds, R., Tolstoy, E., O'Neil Jr., E.J., Hunter, D.A. 1998, AJ, 116, 146
\reference{ } Maeder, A., Conti, P.S. 1994, ARA\&A 32, 227 
\reference{ } Martin, C. 1997, ApJ, 491, 561  
\reference{ } Marzke, R.O., Da Costa, L.N. 1997, AJ, 113, 185
\reference{ } Massey, P. 1998, ApJ, 501, 153
\reference{ } Mateo, M. 1998, ARA\&A, 36, 435
\reference{ } Origlia, L., Leitherer, C. 2000, AJ, submitted
\reference{ } Papaderos, P., Fricke, K.J., Thuan, T.X., Loose, H.-H. 1994,
A\&A, 291, L13
\reference{ } Persson, S. E., Murphy, D.C., Krzeminski, W., Roth, M., Rieke, M.J. 
1998, AJ, 116, 2475
\reference{ } Popescu, C.C., Hopp, U., Hagen, H.J., Els\"{a}sser, H. 1996, A\&AS, 116, 43
\reference{ } Pustil'nik, S., U., Gryumov, A.V., Lipovetsky, V.A., Thuan, T.X., Guseva, N.
1995, ApJ, 443, 499
\reference{ } Salasnich, B., Alessandro, B., Chiosi, C. 1999, A\&A, 342, 131
\reference{ } Salzer, J.J., Moody, W.J., Rosenberg, J.L., Gregory, S.A., Newberry, M.V.
1995, AJ, 109, 2376
\reference{ } Sandage, A. 1986, ARA\&A, 24, 421
\reference{ } Schmidt, A. A., Alloin, D., Bica, E. 1995, MNRAS, 273, 945
\reference{ } Schulte-Ladbeck, R.E., Crone, M.M., Hopp, U. 1998,
ApJ, 493, L23 (SCH98)
\reference{ } Schulte-Ladbeck, R.E., Hopp, U. 1998, AJ, 116, 2886
\reference{ } Schulte-Ladbeck, R.E., Hopp, U., Crone, M.M., Greggio, L. 1999,
ApJ, in press (SHCG99)
\reference{ } Schulte-Ladbeck, R.E., Hopp, U., Crone, M.M., Greggio, L. 2000,
in ``Spectrophotometric Dating of Stars and Galaxies", ed. I. Hubeny, S. Heap \&
R. Cornett (PASP), in press
\reference{ } Searle, L., Sargent, W.L.W. 1972, ApJ, 173, 25
\reference{ } Skillman, E.D. 1989, ApJ, 347, 883
\reference{ } Stetson, P.B., Davis, L.E., Crabtree, D.B. 1990,
in ``CCDs in Astronomy", ed. G.H. Jacoby (PASP), 289
\reference{ } Sung, E.-C., Han, C., Ryden, B.S., Chun, M.-S., Kim, H.-I. 1998,
ApJ, 499, 140
\reference{ } Terlevich, R., Melnick, J., Masegosa, J., Moles, M., Copetti, M.V.F. 1991,
A\&AS, 91, 285
\reference{ } Thuan, T.X. 1983, ApJ, 268, 667
\reference{ } Thuan, T.X. 1985, ApJ, 299, 881
\reference{ } Thuan, T.X., Izotov, Y.I., Lipovetsky, V., Pustilnik, S.A.
1994, in ESO/OHP Workshop on ``Dwarf Galaxies", eds. G. Meylan \& P. Prugniel 
(Garching: ESO), 421
\reference{ } Thuan, T.X., Martin, G.E. 1981, ApJ, 247, 823
\reference{ } Tully, R.B., Boesgaard, A.M., Dyck, H.M., Schempp, W.N. 1981, ApJ,
246, 38
\reference{ } Vanzi, L. 1997, PASP, 109, 1069
\reference{ } Walborn, N., Barb\'{a}, R.H., Brandner, W., Rubio, M., Grebel, E., 
Probst, R.G. 1999, AJ, 117, 225
\reference{ } Worthey, G. 1994, ApJS, 95, 107
\end{references}
\end{document}